\documentclass{article}
\usepackage{spconf,amsmath,epsfig}
\usepackage{graphicx}
\makeatletter
\renewcommand{\maketag@@@}[1]{\hbox{\m@th\normalsize\normalfont#1}}%
\makeatother
\usepackage{caption}
\begin{document}\sloppy

% Example definitions.
% --------------------
\def\x{{\mathbf x}}
\def\L{{\cal L}}

% Title.
% ------
\title{A RETINA-INSPIRED SAMPLING METHOD FOR VISUAL TEXTURE RECONSTRUCTION}
%
% Single address.
% ---------------
\name{Lin Zhu$^{1,\dag}$\thanks{$^{\dag}$ \scriptsize First two authors contributed equally.} \thanks{$^{\ast}$ \scriptsize Corresponding author (E-mail: yhtian@pku.edu.cn).}, Siwei Dong$^{1,\dag}$, Tiejun Huang$^{1,2,3}$, Yonghong Tian$^{1,2,3, \ast}$ \thanks{\scriptsize This work is partially supported by grants from the National Basic Research Program of China under grant 2015CB351806, the National Natural Science Foundation of China under contract No. U1611461, No. 61825101, and No. 61425025.}}

\address{$^{1}$School of EECS, Peking University, Beijing, P.R. China \\$^{2}$School of ECE, Shenzhen Graduate School, Peking University, Shenzhen, P.R. China\\ $^{3}$Pengcheng Laboratory, Shenzhen, P.R. China}

\maketitle

\begin{abstract}
Conventional frame-based camera is not able to meet the demand of rapid reaction for real-time applications, while the emerging dynamic vision sensor (DVS) can realize high speed capturing for moving objects. However, to achieve visual texture reconstruction, DVS need extra information apart from the output spikes. This paper introduces a fovea-like sampling method inspired by the neuron signal processing in retina, which aims at visual texture reconstruction only taking advantage of the properties of spikes. In the proposed method, the pixels independently respond to the luminance changes with temporal asynchronous spikes. Analyzing the arrivals of spikes makes it possible to restore the luminance information, enabling reconstructing the natural scene for visualization. Three decoding methods of spike stream for texture reconstruction are proposed for high-speed motion and stationary scenes. Compared to conventional frame-based camera and DVS, our model can achieve better image quality and higher flexibility, which is capable of changing the way that demanding machine vision applications are built.
\end{abstract}
\section{Introduction}
\label{sec:intro}

Autonomous driving, wearable computing, unmanned aerial vehicles, are typical emerging real-time applications which require rapid reaction in vision processing ~\cite{1}. As the starting point for vision processing, such as foreground detection and object recognition, the step of image sample and texture reconstruction aims to capture and generate the video with high dynamic range and high sensitivity to motion. Thus, the step plays a critical role in practical applications.

\begin{figure}[htb]
 \center\includegraphics[width=7cm]  {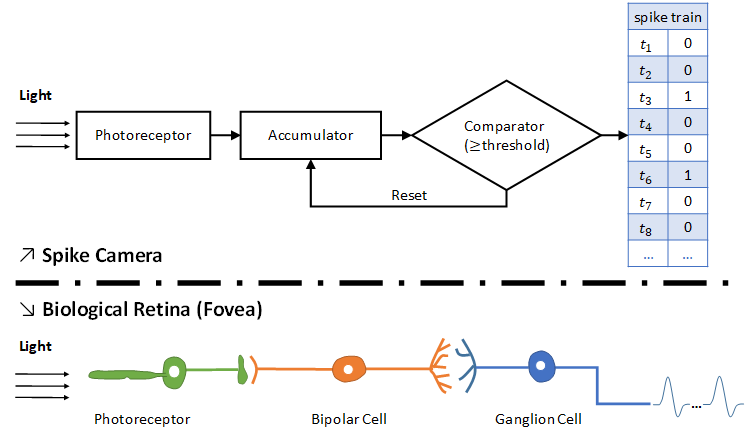}
 \vspace*{-3mm}
 \begin{small}
 \caption{\label{1} The workflow of the spike camera.}
 \end{small}
 \vspace{-0.5cm}
\end{figure}

Modern cameras use CCD (charge-coupled device) or CMOS (complementary metal oxide semiconductor) to capture light and record the motion, which results in a large number of digital pictures and videos ~\cite{2}. Conventional motion pictures are expressed as frame-based images and video sequences. However, the cameras cannot record every moment instantly which leads to discrete simulation as a compromise. The scenes in the exposure time will be compressed into one frame, and motion changes in that time will be blurred. Although there are high speed cameras which are capable of image exposures in excess of 1/1,000 second or frame rates in excess of 250 frames per second ~\cite{3}, the motion changes of high-speed objects in microseconds are missing. For the tasks of high-speed tracking or detection, every moment is important, these consecutive frames have to be compared to recover temporal changes, which is a computationally expensive and difficult task.

Another way is to explore a different sampling method in the direction of biological plausibility. To simulate biological vision, one of the most famous artificial silicon retinas is the dynamic vision sensor (DVS) ~\cite{4}. DVS is capable of detecting and tracking high-speed motion with a high temporal resolution, but it is very difficult to reconstruct the texture, since it only records the change of luminance intensity. Although there are some hybrid sensors combing DVS and conventional image sensor ~\cite{5} or photo-measurement circuit ~\cite{6}, there exists mismatch since the difference of the sampling rate and time.

These bio-inspired approaches concentrate more on modelling the peripheral vision. Nevertheless, the visual detail is acquired through the fovea located in the center of the primate retina, which is responsible for sharp vision (foveal vision) ~\cite{7}, ~\cite{8}. Inspired from the signal processing of neurons, we propose a retina-inspired sampling model and visual texture reconstruction model to address the visual reconstruction problem which is commonly existed in DVS sensors, and integrate them to a novel prototype camera named spike camera. The sampling workflow is shown in Fig. 1.

The rest of the paper is organized as follows: Section 2 briefly reviews related work, and Section 3 analyzes the retina-inspired sampling method. Visual texture reconstruction is presented in Section 4. In Section 5, the experiments on texture reconstruction are conducted. Finally, the paper is concluded in Section 6.
\vspace{-0.3cm}
\section{RELATED WORKS}

\subsection{Dynamic vision sensors}
Dynamic vision sensors (also known as event-based sensors) ~\cite{4} encode the local contrast changes in the scene as positive or negative events at the instant they occur. DVS provides a power efficient way of converting the motion changes into a stream of spatially sparse, temporally dense events. However, the event stream is unsuitable to be directly used as an input to most frame-based computer vision algorithms. To solve this problem, Delbruck et al. developed a bio-inspired DVS camera named DAVIS ~\cite{4}~\cite{5}. DAVIS integrates DVS with a frame-based active-pixel sensor (APS). This makes it able to capture frame-based texture. However, the difference of the sampling rate between APS (60 frames per second) and DVS (a temporal resolution of 1 $\mu s$), the mismatch is quite obvious. Recently, Posch et al. ~\cite{6} tried a different way by introducing the asynchronous time-based image sensor (ATIS). ATIS consists of a DVS circuit and a photo-measurement circuit. The photo-measurement is triggered while the DVS circuit fires a spike (i.e. intensity change detected), and the time of this process is encoding. Since the intensity is inversely proportional to the integration time, the visual texture can be reconstructed. ATIS seems to be perfect, but the intensity is only measured when a DVS spike is generated, and the measurement is posterior to the spike firing, so the mismatch still exists, especially in flexible motion.
\vspace{-0.2cm}
\subsection{The integrate-and-fire model}

The fovea is responsible for the sharpest vision in primates eyes. In the fovea centralis of the primate retina, the bipolar cell connects a single cone cell (photoreceptor) with a single ganglion cell, which provides the sharpest vision ~\cite{7}. If a photon is captured by the photoreceptor, the optical-to-electrical conversion is triggered. The electrical signal is transferred to bipolar cell and ganglion cell. With more photons collected, the membrane potential of the ganglion cell hits the threshold, a spike is generated. Then the membrane potential decays back to the resting potential ~\cite{8}. The structure of fovea centralis and the signal processing have stimulated the inspiration of the development of retinomorphic vision sensors. This procedure is in accordance with the integrate-and-fire model.

The integrate-and-fire (IF) model ~\cite{9} is one of the most commonly used neuron models in neuroscience. In the IF model, the neuron is considered as a leaky capacitor. When the stimuli current is conducted, the membrane is charged. Once the membrane voltage reaches the threshold, a spike is fired. After that the membrane potential is reset to the resting potential. Compared to a real spiking neuron, the integrate-and-fire model is fairly simple. However, in neuromorphic engineering, the model has been widely used in building sensory systems. Another representative neuron model is the spike response model (SRM), which is a generalization of the leaky integrate-and-fire model ~\cite{10}. The neuron model is interpreted as a membrane filter and a function describing the shape of the spike. In contrast to the integrate-and-fire model, SRM is somewhat simpler on the level of the spike generation mechanism. But the threshold behavior of the SRM is richer than integrate-and-fire model, it can account for various aspects of refractoriness and adaptation.

\vspace{-0.2cm}

\section{RETINA-INSPIRED SAMPLING METHOD}

\subsection{Retina-inspired sampling method}

In fovea, a bipolar cell only contacts one photoreceptor and one ganglion cell. To a first and rough approximation, neuronal dynamics can be conceived as a summation process (sometimes also called `integration' process) combined with a mechanism that triggers action potentials above some critical voltage. Inspired by the above, we propose a novel sampling model and integrate it into a prototype camera called spike camera.

In spike camera, the intensity of light is converted into voltage by the photoreceptor. Once the analog-to-digital converter (ADC) completes the signal conversion and outputs the digital luminance intensity, the accumulator at each pixel accumulates the intensity. Different luminance intensities lead to different accumulate rates. For a pixel, if the accumulated intensity reaches the dispatch threshold $\phi$, a spike is fired indicates that the luminance here is large enough (as Eq. (1)). At the same time, the corresponding accumulator is reset in which all the charges are drained.
{\setlength\abovedisplayskip{1pt}
{\setlength\belowdisplayskip{1pt}
\begin{eqnarray}
\int_{0}^{t}Idt\geq \phi
\end{eqnarray}
where $I$ refers to the luminance intensity and $t$ indicates the integration time.

Typically, we can build a pixel array and in each pixel a photoreceptor and its corresponding accumulator are set. The output and the reset are triggered asynchronously. Here at each sampling moment, if a spike is just fired, a digital signal ``1" is outputted, otherwise ``0" is generated.
\begin{figure}[htb]
 \center{\includegraphics[width=7cm]  {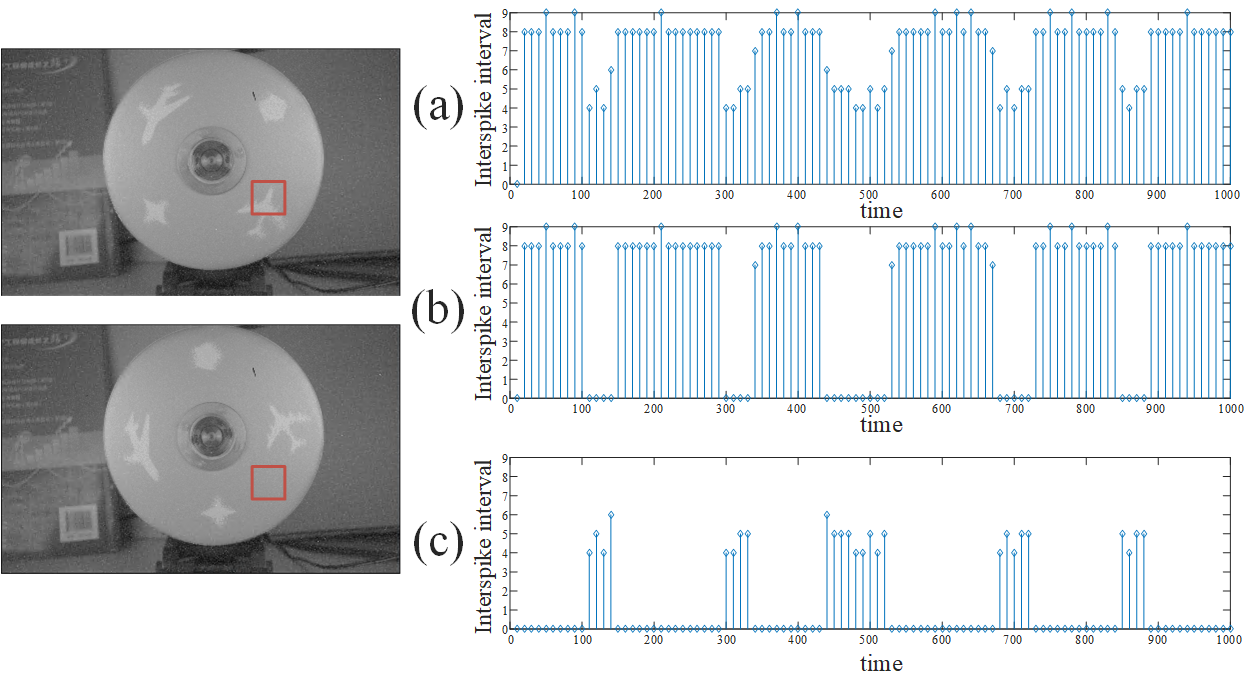}}
 \begin{small}
 \caption{\label{2} The temporal distribution of spike train. (a) The ISI distribution in temporal domain; (b) The temporal distribution corresponding to dark object (the disc); (c) The temporal distribution corresponding to bright object (the patterns of plane and pentagram).}
 \end{small}
 \vspace{-0.3cm}
 \end{figure}
Compared to the conventional camera, the spike camera only cares about the luminance intensity. Considering that at different pixel, the accumulating speed of the luminance intensity is quite different; the patterns of dispatched spike train are as well different from each other. For brighter pixels, ``1" appears more frequently than that of darker ones. The idea is easy to explain. The brighter pixel indicates that more photons are collected at the pixel resulting in larger ADC value which is easier and faster to exceed the dispatch threshold. According to this principle, the texture can be reconstructed by analyzing the patterns of spikes. It is quite similar to the processing of the responses of ganglion cell which illustrates the outline of the object by decoding the spike latencies ~\cite{11}.
\vspace{-0.2cm}
\subsection{The temporal correlation of spike train}

The spike camera is able to record spike data with microsecond-level resolution. Theoretically, we can get the luminance intensity information at arbitrary time from the spike. This characteristic enables it to be applied to a variety of vision tasks, such as object detection, tracking. Also, the high temporal correlation between each consecutive spike makes it is benefit to spike coding and compression.

Assuming that the luminance intensity is a constant in a short period. Based on the mechanism of spike generation, Eq. (1) can be simplified as $\bar{I}\Delta t\geq \phi$, where $\Delta t$ is the inter-spike interval (ISI) obtained by calculating the time between two neighboring spikes. Therefore, the average intensity of the pixel in this period can be estimated by
\begin{eqnarray}
\bar{I}=\frac{\phi}{\Delta t}
\vspace{-0.7cm}
\end{eqnarray}

In order to analyze the spike train intuitively, we give the distribution of the ISI in the temporal domain. As shown in Fig. 2, the video depicts a spinning disc with 2000 revolutions per minute (rpm). We select a pixel located in the red box (as shown in the left picture in Fig. 2). The patterns on the disc are brighter, thus their corresponding intervals are smaller than that of the disk. The results show clearly that the spike train has a high correlation in the temporal domain. We can easily distinguish patterns and disk by a simple threshold or other probability-based models.

\vspace{-0.25cm}
\section{VISUAL TEXTURE RECONSTRUCTION}

To restore the captured scene and bridge the gap between the asynchronous bionic spike data and conventional frame-based vision, we propose several visual texture reconstruction strategies aiming at various tasks and applications, such as high-speed motion and stationary scenarios. The spikes can be utilized according to two principles: 1) the intensity is inversely proportional to the ISI; 2) the intensity is directly proportional to the spike counts or spike frequency. Thus, by taking advantage of the ISIs or simply counting the spikes over a period, the scene is able to be fully reconstructed.
\begin{figure}[t]
 \center{\includegraphics[width=7cm]  {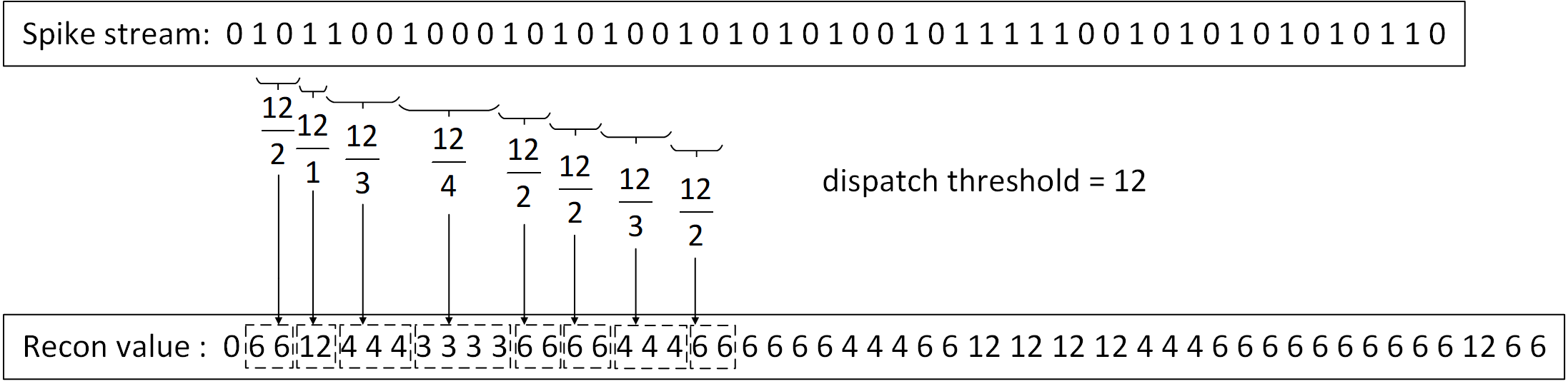}}
 \begin{small}
 \caption{\label{3} The texture reconstruction from ISI (TFI).}
 \end{small}
 \end{figure}
 \begin{figure}[t]
 \center{\includegraphics[width=7cm]  {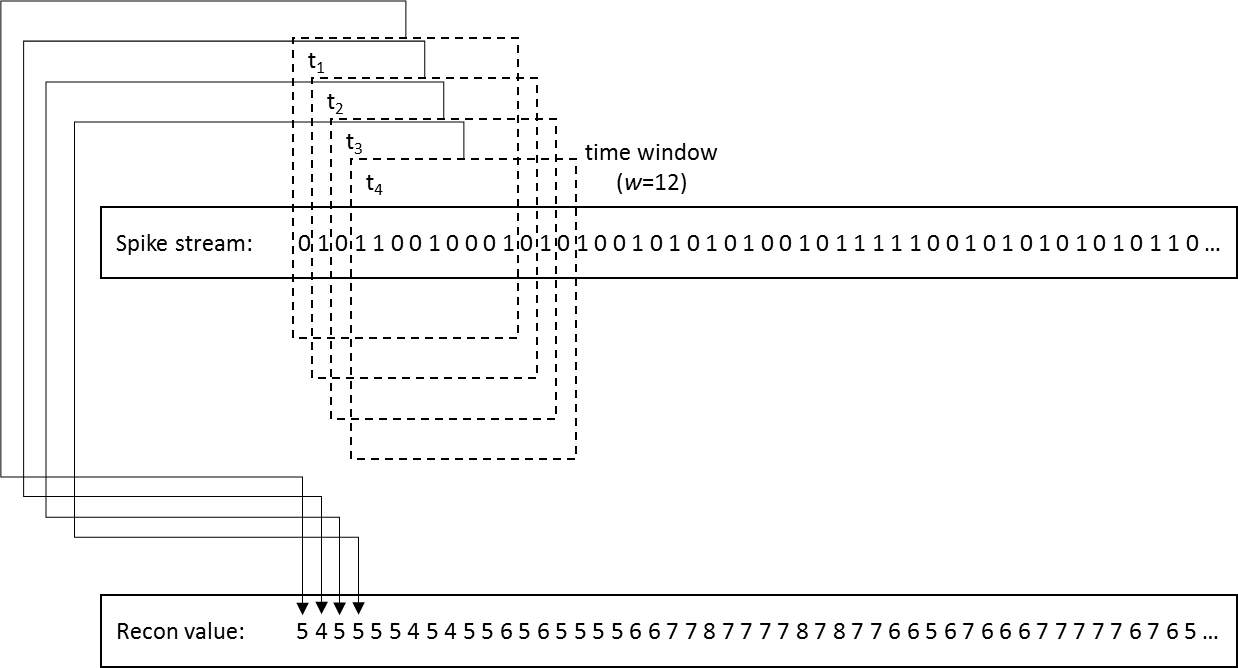}}
 \begin{small}
 \caption{\label{4}  The texture reconstruction from playback with the moving time window (TFP).}
 \end{small}
 \end{figure}
\vspace{-0.2cm}
\subsection{Texture reconstruction for high-speed motion}

For high-speed motion applications, the spike firing patterns are rapidly changing. From the first principle mentioned above, the texture is reconstructed from the ISIs, shown in Fig. 3. According to Eq. (2), the reconstructed the pixel value can be estimated by the mean luminance intensity using only two spikes (i.e. one ISI).
\begin{eqnarray}
P_{t_{i}}=\frac{C }{\Delta t_{i}}
\vspace{-0.5cm}
\end{eqnarray}
where $P_{t_{i}}$ refers to the pixel value at the moment of $t_{i}$, $C$ denotes the maximum dynamic range, and $\Delta_{t_{i}}$ means the ISI between $t_{i}$ and the last moment when spike appears. Compared to the original texture, the reconstructed values may not be perfectly matched. But the variation is in accordance with the original one. This method (Texture from ISI, TFI) can reconstruct the outline of the texture but not the clear details. When the object moves very quickly, the picture reconstructed from luminance intensity performs the motion nearly synchronous.
 \vspace{-0.4cm}
\subsection{Texture reconstruction for stationary scenes}

For stationary scenes, the spike firing characteristics are rarely changed. By taking advantage of the second principle, if we play back the spikes, the historical pictures are able to be illustrated. In this method (Texture from Playback, TFP), there is a moving time window collecting the spikes in a specific period. By counting these spikes, the texture is computed as:
\begin{eqnarray}
P_{t_{i}}=\frac{N_{w}}{w}\cdot C
\vspace{-0.38cm}
\end{eqnarray}

As shown in Fig. 4, the size of the time window is $w$ which refers to the previous $w$ moments before $t_{i}$. $N_{w}$ is the total number of spikes collected in the time window. $C$ refers to the maximum dynamic range of the reconstruction. The larger size means the playback period is longer and more spikes are included. When the time window size is set to the dispatch threshold $\phi$, the textures are accurately reconstructed. Meanwhile, the TFP method could restore the texture with various dynamic ranges by resizing the time window to the value of different contrast levels ~\cite{12} ~\cite{13}.

 \vspace{-0.2cm}
\subsection{Adaptive texture reconstruction in a bio-inspired way}

TFP method reconstructs texture by playing back the spikes in a historical window, which incurs a trade-off between length of time-window and latency. If the texture can be reconstructed adaptively according to the influence of historical spikes, instead of a predefined window, better results will be achieved. On the other hand, in biological retina, the spikes are conveyed to the visual cortex for advanced analysis in which various neurons process the inputs and response with their own firing characteristics. For simplicity, these neurons can be abstracted as a typical neuron but with different membrane potential thresholds to adapt to different stimuli. To this end, an adaptive texture reconstruction method (Texture from Adaptive threshold, TFA) based on SRM is proposed.

The state of a neuron is described by a single variable $u$ which we interpret as the membrane potential. In the absence of input, the variable $u$ is at its resting value ($u_{rest}=0$ in this work). A short current pulse will perturb $u$ and it takes some time before $u$ returns to rest.

If we regard each pixel as a neuron, the output of spike camera can be considered as the input current $I\left ( t \right )$ of the model, which is a unit step function. First, the input current  $I\left ( t \right )$  is filtered with a filter $\kappa \left ( s \right )$ and yields the input potential $h\left ( t \right )$.
\begin{eqnarray}
h\left ( t \right )=\int_{0}^{t}\kappa _{t}\left ( s \right )I\left ( s \right )ds
\end{eqnarray}
where $\kappa \left ( s \right )$  describes the time course of the voltage response to a short current pulse at time $s$. It is also a function describing the shape of the potential, according to SRM model [9],$\kappa \left ( s \right )$ is defined as:
\begin{eqnarray}
\kappa _{t}\left ( s \right )=\frac{t-s-\Delta }{\tau }\exp\left ( 1-\frac{t-s-\Delta }{\tau } \right )
\end{eqnarray}
where $\Delta$ and $\tau$ are the parameters to adjust the shape of the model. In general, $\Delta$ is set to zero, $\tau$ is a constant.
\begin{figure}[htb]
 \center{\includegraphics[width=8.5cm]  {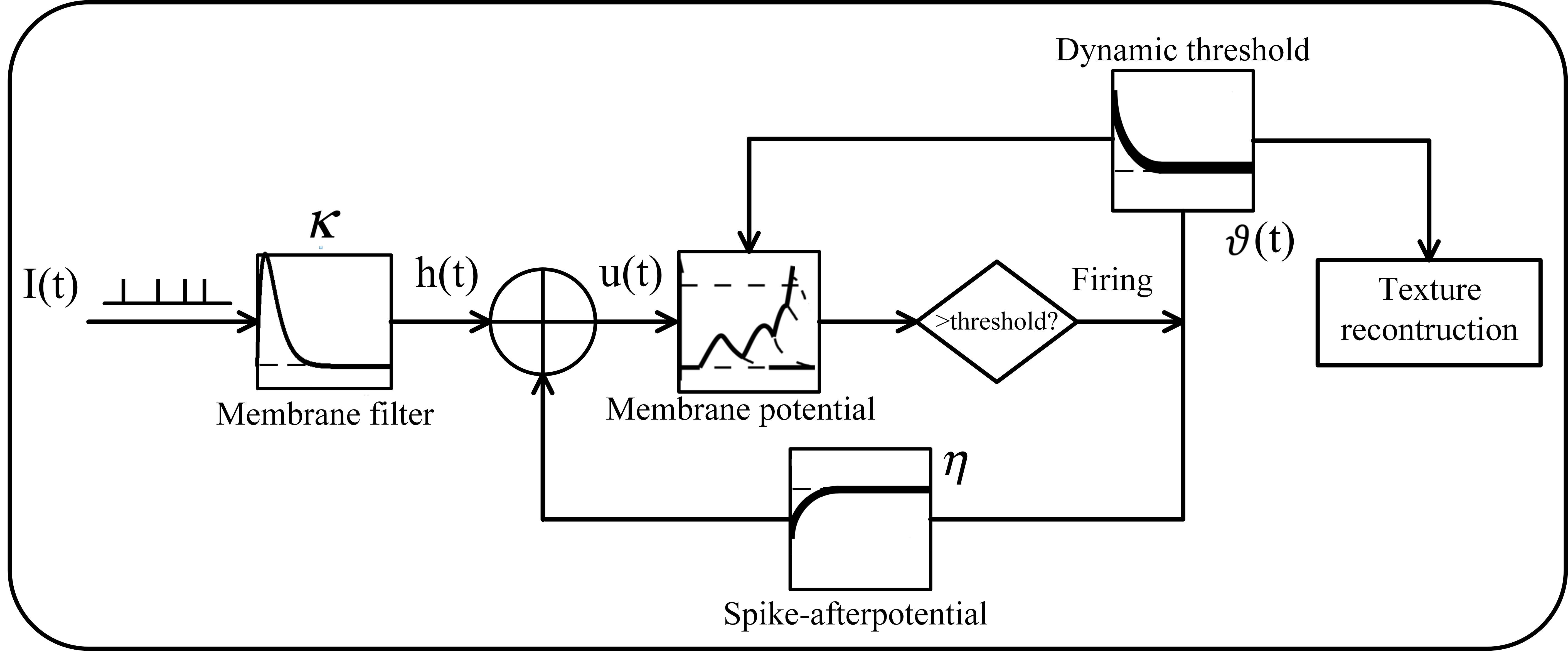}}
 \begin{small}
 \caption{\label{5}  The workflow of adaptive texture reconstruction (TFA).The spike train is considered as the input current $I \left ( t \right )$ of neuron. It is filtered with $\kappa \left ( s \right )$ and yields a potential $h\left ( t \right )$. Firing occurs if the accumulated membrane potential $u \left ( t \right )$ reaches the threshold $\vartheta \left ( t \right )$. Then the model is adaptively adjusted to fit the input current, by dynamic threshold and spike-afterpotential. The dynamic threshold $\vartheta \left( t \right )$ is a learning process for the feature of input current, thus it is suitable for describing the texture.}
 \end{small}
\vspace{-0.4cm}
 \end{figure}
\begin{figure}[htb]
 \center{\includegraphics[width=8.5cm]  {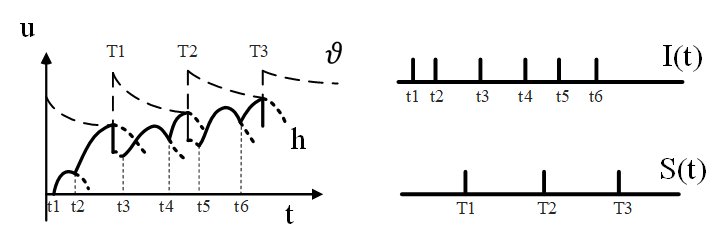}}
 \vspace*{-3mm}
 \begin{small}
 \caption{\label{6}  The adaptive dynamic threshold. Input current $I_{t}$ yields the input potential. At time $T$ a spike occurs because the membrane potential hits the threshold $\vartheta \left ( t \right )$. The threshold jumps to a higher value (dashed line) and, at the same time, a decay is added to the membrane potential. If no further spikes are triggered, the threshold decays back to its resting value.}
 \end{small}
 \vspace{-0.2cm}
 \end{figure}

Then, the evolution of $u$ is given by
\begin{eqnarray}
u\left ( t \right )=\int_{0}^{\infty }\eta \left ( z \right )S\left ( t-z \right )dz+h\left ( t \right )+u_{rest}
\end{eqnarray}
where $\eta \left ( \cdot  \right )$ is the voltage contribution after firing a spike, which feeds back to the membrane potential. We set $\eta \left ( s  \right ) = 1$ in experiment.
\begin{figure*}[htb]
 \center{\includegraphics[width=16cm]  {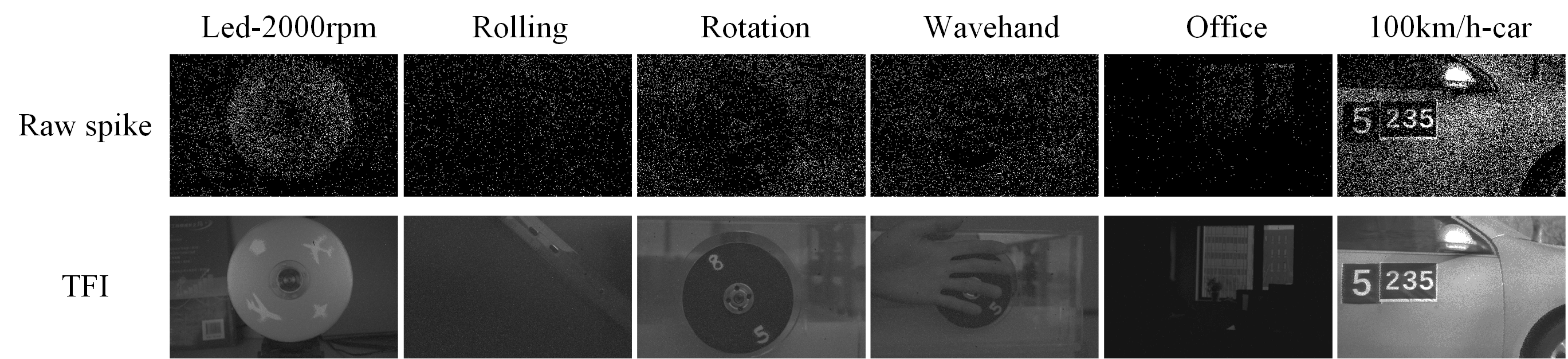}}
 \vspace*{-3mm}
 \caption{\label{7}  Texture reconstruction via TFI.}
 \end{figure*}
\begin{figure*}[htb]
 \center{\includegraphics[width=16cm]  {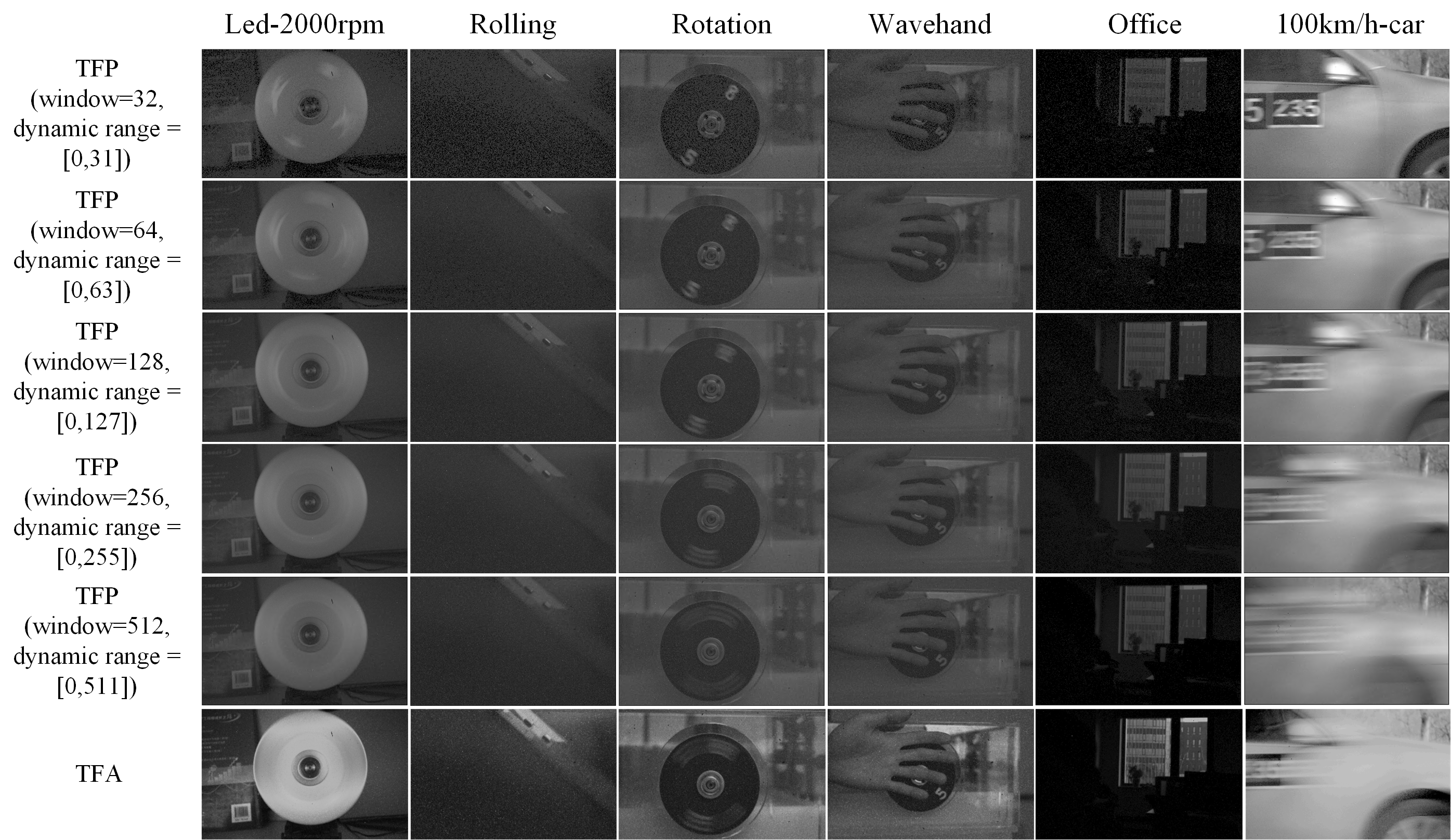}}
 \vspace*{-3mm}
 \caption{\label{8}  Texture reconstruction via TFP and TFA.}
 \end{figure*}
Spike firing is defined by a threshold process. If the membrane potential $u$ reaches the threshold $\vartheta$, an output spike is triggered. Meanwhile, a dynamic adjusting is performed for threshold to adaptive the firing rate. The model of the dynamic threshold is:
\begin{eqnarray}
\vartheta \left ( t_{i} \right )=\vartheta \left ( t_{i-1} \right )+f_{t_{i}}\left ( s \right )
\end{eqnarray}
where
\begin{small}
\begin{eqnarray}
f_{t_{i}}\left ( s \right )=\left\{\begin{matrix}
\exp \left ( -\int_{0}^{T} S\left ( t-s \right )ds\right )\: \: \:  if\: spike\: firing\: at\: t_{i}\\
-\exp \left ( -2*\int_{0}^{T}S\left ( t-s \right )ds \right )\:\: \:  otherwise\: \: \: \: \: \: \: \: \: \:
\end{matrix}\right.
\end{eqnarray}
\end{small}
where $\vartheta \left ( t_{0} \right )$ is the initial threshold, $f_{t_{i}} \left ( s \right )$ is an adaptive function to adjust the threshold, and $T$ is time range for calculating the number of spikes. The threshold is adjusted by the amount of firing times in the past $T$ time.

The spike firing threshold reflects the intensity of external stimuli, it is the adaptation of the neuron model to the environment. Threshold adjustment can be seen as a learning process of model to the firing rate. At a certain time, if we put the thresholds of all the pixels together and form a matrix, the visual texture reconstruction is done naturally after normalization.

 \vspace{-0.1cm}

\subsection{Motion representation}

Dynamic vision sensors care about the intensity changes, thus they only response with events in the motion regions. In contrast, the spike camera collects the full-time luminance intensity information. As a consequence, the spike data can be theoretically converted into the events of DVS, which consist of the timestamp, pixel-location and the polarity of intensity change (On or Off).

In DVS, an event is triggered once the change in log intensity at a pixel exceeds a preset threshold.  If the last event generated by DVS occurs at luminance intensity $I_{\mu}$, a new event is fired at $I_{\nu}$ when a change in that log intensity exceeds threshold $\theta$:
\begin{eqnarray}
\left |\log\left ( I_{\mu} \right )- \log\left ( I_{\nu} \right )  \right |\geq \theta
\end{eqnarray}

In spike camera, $I_{\mu}$ and $I_{\nu}$ can be estimated by Eq. (2). Then, we have
\begin{eqnarray}
\left |\log\left ( \frac{\phi }{\Delta t_{\mu}} \right )- \log\left ( \frac{\phi }{\Delta t_{\nu}} \right )  \right |\geq \theta
\end{eqnarray}

Thus, Eq. (11) can be simplified as:
\begin{eqnarray}
\left |\log\left ( \Delta t_{\mu} \right )- \log\left (\Delta t_{\nu} \right )  \right |\geq \theta
\end{eqnarray}
where $\Delta t$ is the ISI of the spike data. Consequently, the spike data can be converted into events like that of DVS. This is one of the advantages of the proposed sampling method. The corresponding experiments are conducted in Section 5.

\section{EXPERIMENTS}

\subsection{Experimental setup}

We test proposed visual texture reconstruction methods on spike sequences captured by the spike camera. The details of each spike sequences are shown in Table I.

 \vspace{-0.3cm}
\subsection{Visual texture reconstruction}

One of the most important applications of the spike camera is for visual-friendly viewing. The texture reconstruction experiment is conducted utilizing three methods: TFI, TFP and TFA. To obtain visualization images, the gamma correction $\left ( \gamma =2.2 \right )$ is applied to the reconstructed textures of TFI and TFP, while TFA shows the original results.

The experimental results of TFI are shown in Fig. 7. The result indicates that the outline of the object is quite clear, while some detailed textures are missing. Thus, TFI is more suitable for real-time applications and some detection related tasks.
\newcommand{\tabincell}[2]{\begin{tabular}{@{}#1@{}}#2\end{tabular}}
\begin{table}[t]
\small
	\begin{center}
		\caption{Information of the spike sequences.} \label{tab:cap}
 \vspace{-0.2cm}
		\setlength{\tabcolsep}{1mm}{
			\begin{tabular}{|c|c|c|c|}
				\hline
				\tabincell{c}{\textbf{Spike}\\ \textbf{sequences}} & \textbf{Size$\times$length} & \textbf{Sampling rate} & \textbf{Description}
				\\
				\hline
				rotation  & $\left [ 400 , 250 \right ]\times153600$ & 40000Hz & High-speed \\
				wavehand & $\left [ 400 , 250 \right ]\times153600$ & 40000Hz & Normal-speed \\
				rolling & $\left [ 400 , 250 \right ]\times153600$ & 40000Hz & Normal-speed \\
				office & $\left [ 400 , 250 \right ]\times153600$ & 40000Hz & Static \\
                led-rpm2000 & $\left [ 400 , 250 \right ]\times200000$ & 20000Hz & High-speed \\
                100km/h-car & $\left [ 400 , 250 \right ]\times200000$ & 20000Hz & High-speed \\
				\hline
				
			\end{tabular}
		}
	\end{center}
\vspace{-0.7cm}
\end{table}

For static scenes, TFP and TFA reconstruct texture with more details and higher dynamic range, as shown in Fig. 8. For TFP method, the size of time window is a key parameter in texture reconstruction. We select five typical sizes for comparison which are 32, 64, 128, 256 and 512. From the results, the static region is reconstructed with more details when higher dynamic range is gained. Smaller time window size achieves clearer outline for moving objects but lower quality for static background pixels.

TFA can be considered as an adaptive version of TFP, the results achieve better performance than TFP. The high-speed motion of the object is blurred in these two methods, but the texture details are reconstructed more completely. The threshold adjustment of TFA is shown in Fig. 9. For a static pixel, the threshold converges to a constant value after about 500 moments. For the sample rate of 40000Hz, it needs about 0.0125s to adjust the model.

The spike train is able to be converted into the events of DVS according to Eq. (12). The experimental results are shown in Fig. 10, which shows that the events are implicit in spike data. Thus, the spike data is suitable to achieve the motion detection and object tracking tasks.

 \vspace{-0.3cm}
\section{CONCLUSION}

In this paper, a novel bio-inspired spike camera is introduced which simulates the retinal imaging. Three decoding methods of spike train for texture reconstruction are proposed which enables playing back any historical moment. Experimental results show that TFI is more suitable for real-time applications such as object detection or action recognition related tasks, TFP is capable of reconstructing visual-friendly picture sequences for viewing, while TFA is adaptive and give a high-quality texture in static scenes. For future work, more efficient texture reconstruction methods need to be explored, especially in complex scenes.
 \begin{figure}[htb]
 \center{\includegraphics[width=6cm]  {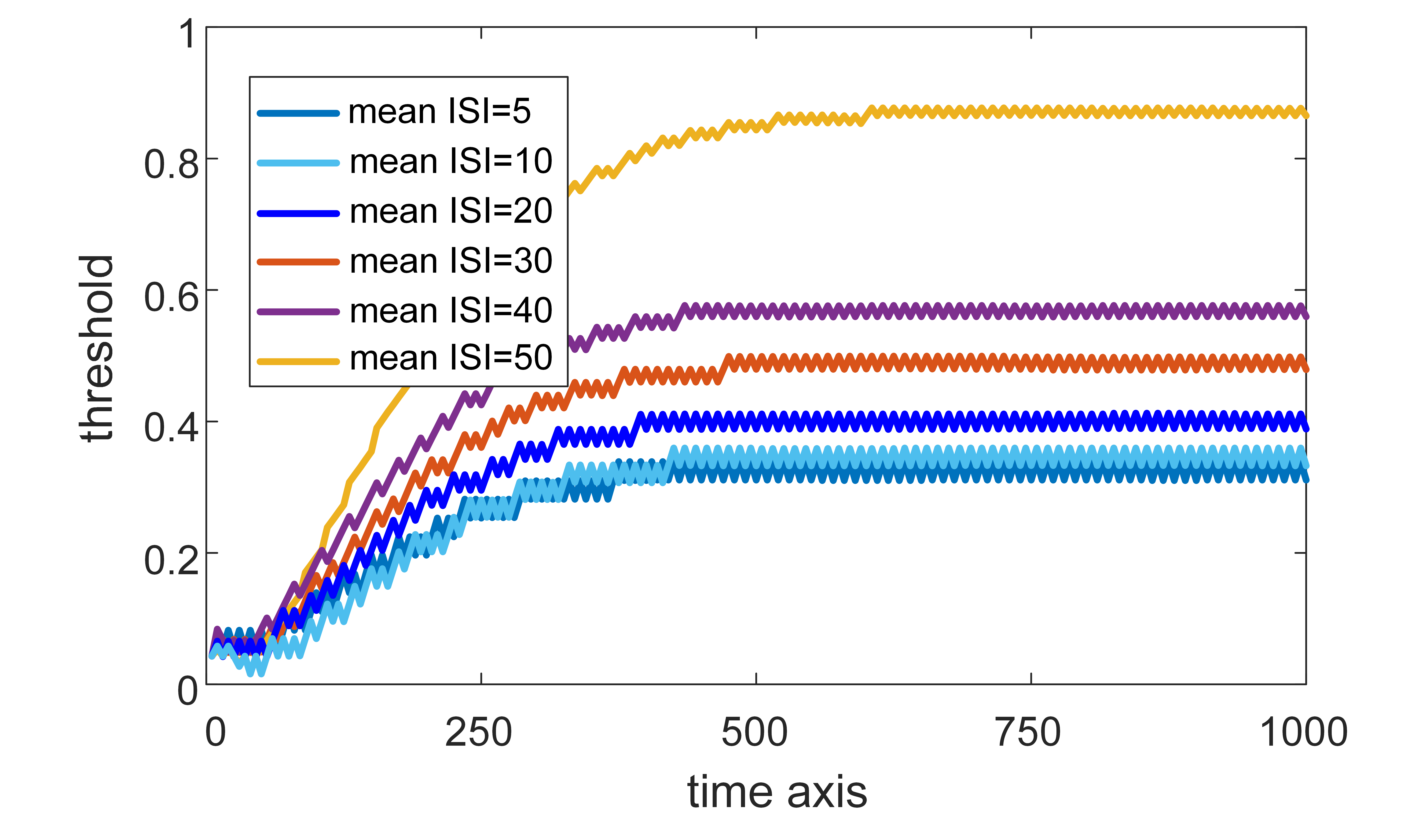}}
 \vspace*{-3mm}
 \begin{small}
 \caption{\label{9}  The dynamic threshold in TFA.}
 \end{small}
 \end{figure}
 
 \begin{figure}[htb]
 \center{\includegraphics[width=8cm]  {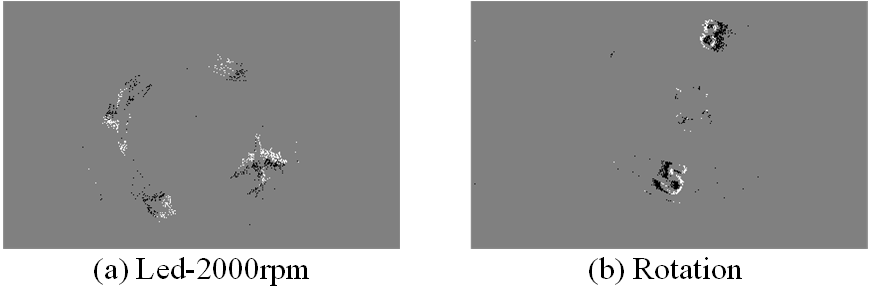}}
 \vspace*{-3mm}
 \begin{small}
 \caption{\label{10}  Convert the spike data to events. White pixels denote ``On" events and black ones denote ``Off" events.}
 \end{small}
 \end{figure}

\linespread{0.3}
\small
\bibliographystyle{IEEEbib}
\bibliography{icme2019template}

\begin{thebibliography}{10}

\bibitem{1}
A.~Hilton T.~Moeslund and V.~Kruger,
\newblock ``A survey of advances in vision-based human motion capture and
  analysis,''
\newblock {\em Computer vision and image understanding}, 2006.

\bibitem{2}
G.~C. Holst and T.~S. Lomheim,
\newblock ``Cmos/ccd sensors and aamera systems,''
\newblock {\em Bellingham, WA: SPIE}, 2007.

\bibitem{3}
A.~Hilton T.~Moeslund and V.~Kruger,
\newblock ``High-speed camera observations of negative ground flashes on a
  millisecond-scale,''
\newblock {\em Geophys. Res. Lett.}, vol. 32, no. 23, pp. L23802, 2005.

\bibitem{4}
C.~Posch P.~Lichtsteiner and T.~Delbruck,
\newblock ``A 128$\times$128 120db 15$\mu$s latency asynchronous temporal
  contrast vision sensor,''
\newblock {\em IEEE J. Solid-st. Circ.}, vol. 43, no. 2, pp. 566--576, 2008.

\bibitem{5}
M.~Yang S.~Liu C.~Brandli, R.~Berner and T.~Delbruck,
\newblock ``A 240$\times$180 130db 3$\mu$s latency global shutter
  spatiotemporal vision sensor,''
\newblock {\em IEEE J. Solid-st. Circ.}, vol. 49, no. 10, pp. 2333--2341, 2014.

\bibitem{6}
D.~Matolin C.~Posch and R.~Wohlgenannt,
\newblock ``An asynchronous time-based image sensor,''
\newblock {\em IEEE International Symposium on Circuits and Systems}, pp.
  2130--2133, 2000.

\bibitem{7}
A.~Kaehler et~al. S.~Gould, J.~Arfvidsson,
\newblock ``Peripheral-foveal vision for real-time object recognition and
  tracking in video,''
\newblock {\em IJCAI}, 2007.

\bibitem{8}
F.~Arrebola et~al. R.~Marfil, E.~Antunez,
\newblock ``Merging attention and segmentation: active foveal image
  representation,''
\newblock {\em Brain-Inspired Computing,Springer International Publishing},
  2013.

\bibitem{9}
A.~N. Burkitt,
\newblock ``A review of the integrate-and-fire neuron model: I. homogeneous
  synaptic input,''
\newblock {\em Biol. Cybern.}, vol. 95, pp. 1--19, 2006.

\bibitem{10}
M.~Avermann et~al. S.~Mensi, R.~Naud,
\newblock ``Parameter extraction and classification of three neuron types
  reveals two different adaptation mechanisms,''
\newblock {\em J. Neurophys.}, vol. 107, pp. 1756--1775, 2007.

\bibitem{11}
T.~Gollisch and M.~Markus,
\newblock ``Rapid neural coding in the retina with relative spike latencies,''
\newblock {\em Science}, pp. 1108--1111, 2008.

\bibitem{12}
W.~Heidrich H.~Seetzen and et~al. W.~Stuerzlinger,
\newblock ``High dynamic range display systems,''
\newblock {\em ACM SIGGRAPH}, pp. 760--768, 2004.

\bibitem{13}
S.Winder et~al. S.B.Kang, M.~Uyttendaele,
\newblock ``High dynamic range video,''
\newblock {\em ACM Transactions on Graphics}, vol. 22, no. 3, pp. 319--325,
  2003.

\end{thebibliography}

\end{document}